**Magnetic domains without domain walls: a unique effect of He$^+$ ion bombardment in ferrimagnetic Co/Tb multilayers.**

Łukasz Frąckowiak[1], Piotr Kuświk[1], Gabriel David Chaves-O'Flynn[1], Maciej Urbaniak[1], Michał Matczak[2], Andrzej Maziewski[2], Meike Reginka[3], Arno Ehresmann[3], and Feliks Stobiecki[1]

[1] Institute of Molecular Physics, Polish Academy of Sciences, Poznań, Poland
[2] Faculty of Physics, University of Białystok, Białystok, Poland
[3] Institute of Physics and Center for Interdisciplinary Nanostructure Science and Technology (CINSaT), University of Kassel, Kassel, Germany

**Abstract**

We show that it is possible to engineer magnetic multi-domain configurations without domain walls in a prototypical rare earth/transition metal ferrimagnet using keV He$^+$ ion bombardment. We additionally shown that these patterns display a particularly stable magnetic configuration due to a deep minimum in the free energy of the system which is caused by flux closure and the corresponding reduction of the magnetostatic part of the total free energy. This is possible because light-ion bombardment differently affects an elements relative contribution to the effective properties of the ferrimagnet. The impact of bombardment is stronger for rare earth elements. Therefore, it is possible to influence the relative contributions of the two magnetic subsystems in a controlled manner. The selection of material system and the use of light-ion bombardment open a route to engineer domain patterns in continuous magnetic films much smaller than what is currently considered possible.

**Introduction**

The ability to create lateral magnetic domain patterns is at the heart of a manifold of applications. Their use in magnetic mass memories [1–4] is absolutely straightforward but also in other areas, such as magnonics [5,6], or for the formation of defined domain patterns used for magnetophoresis in lab-on-a-chip devices [7–11] magnetic domain engineering forms a basic technology. For such applications, it is common to use ferromagnetic layers. In these materials, magnetic domains are uniformly magnetized regions, in which the effective magnetization points in a definite direction. Naturally occurring domain patterns are formed by free energy minimization, usually as a compromise between exchange, anisotropy and stray field energy terms [5,7,12,13]. The domains are separated by domain walls (DWs) which are the transition regions where the magnetic moment reorients from the direction within the first domain to the direction within the second. DW geometries depend on the ratio between the exchange coupling and the anisotropy constants and typically consist of a narrow core and comparably wide tails [12]. Their widths constitute the natural size limit for individual domains. Therefore, the lateral DW widths also constitute the critical dimension for magnetic domain engineering in continuous layers. Domain patterns can be engineered by local modification of magnetic properties such as the coercive field ($H_C$) [14–17] or the exchange bias coupling of systems composed of ferromagnetic and antiferromagnetic layers [18–21]. In the past this has been achieved, e.g., by light-ion bombardment through masks [18–20,22,23], by focused ion beams [24–27], by direct laser writing [28], or by thermally assisted scanning probe lithography [29]. Walls between magnetic domains engineered by these methods are usually non-symmetric [30] with respect to their center due to different anisotropies on the two sides of the wall. However, even those methods will not be able to engineer domains of lateral dimensions below the respective (average) DW widths. Here we describe a ferrimagnetic material system in combination with a method to engineer magnetic domain patterns without lateral DWs. This unique combination promises magnetic domains in continuous layer systems of dimensions well below the typical ferromagnetic DW widths.

The fundamental physics of magnetic domain formation in ferrimagnetic films is similar to the one in ferromagnetic films [12], the occurrence of two magnetic moment subsystems, however, results in more involved domain formation effects. Layer systems consisting of rare earth (RE) transition metal (TM) alloyed layers, with alternating stoichiometric domination of RE (RE+) or TM (TM+) will contain interfacial DWs at saturation [31–37] (Fig.1a). This peculiar situation is possible because parallel effective magnetizations (black arrows in Fig. 1) in the RE+ and TM+ layers correspond to antiparallel magnetic moments of the magnetic subsystems (red and blue arrows in Fig. 1) of the same type (RE or TM) in the two homogeneous different layers [34]. Recently Li and coworkers investigated RE-TM alloy films with inhomogeneous concentrations of Tb [38] [Li2016] whose magnetization reversal characteristics have also been explained by the existence of two nanoscale amorphous phases in a TbFeCo film with differing Tb concentration.

Here, we demonstrate that 10 keV $He^+$ ion bombardment allows to modify the magnetic properties of ferrimagnetic Co/Tb multilayers that exhibit perpendicular magnetic anisotropy [39–43]. In particular, we show that with increasing dose of $He^+$ ions the Tb magnetization decreases much stronger than the Co one. This finding opens a way to pattern RE+ ferrimagnetic films by light-ion bombardment through a mask or by light-ion beam writing to locally reverse the domination from RE+ to TM+ and therefore engineer magnetic domains without DWs in the two magnetic subsystems (Fig. 1b). Using this patterning technique, we fabricate a laterally periodic domain pattern consisting of a lattice of low $H_C$ TM+ squares embedded in a high $H_C$ RE+ grid (later referred to as matrix).

**Results and discussion**

The subjects of our investigations are Tb/Co multilayers displaying, for small sublayer thicknesses, magnetic properties similar to amorphous Co-Tb alloy films [39–43]. In order to determine the influence of the 10 keV $He^+$ ion bombardment on the properties of the $(Tb/Co)_6$ multilayers as a function of the thickness ratio between Co and Tb layers, i.e. as a function of the effective multilayer composition a particular layer system was deposited. In this layer system, the nominal thicknesses of the Co sublayers were fixed at $t_{Co} = 0.66$ nm and the Tb sublayers were deposited as wedges with thicknesses $0 \leq t_{Tb} \leq 2$ nm. The sample was bombarded with the two different $He^+$-ion doses $D$ of $1\times10^{15}$ and $3\times10^{15}$ ions/cm² (see description in methods). The characterization of magnetic properties was performed using a Magnetooptical Kerr Effect (MOKE) magnetometer in polar configuration with a probing-light wavelength of 640 nm. Fig. 2 shows changes of the coercive field as a function of the Tb sublayer thickness ($H_C(t_{Tb})$) for an unbombarded area and two areas bombarded with $D = 1\times10^{15}$ $He^+/cm^2$ and $D = 3\times10^{15}$ $He^+/cm^2$. The singularities in the curves $H_C(t_{Tb};D)$ correspond to the Tb layer thicknesses $t_{Tb}$ and the associated effective Tb concentration $c_{Tb}$ at which the magnetic moments of Co and Tb compensate each other. It is easily seen that these values increase with increasing $D$.

Note that the hysteresis loops for systems with Tb and Co domination have opposite orientations. This occurs because for the light wavelength used in the MOKE set up the Co magnetic subsystem determines the sign of the magnetooptical signal; the Co magnetic moments are parallel to the net magnetization in Co dominated films, and antiparallel in Tb dominated films [39,44–46]. After ion bombardment with $D = 1\times10^{15}$ $He^+/cm^2$ (Fig. 2c), the hysteresis loop still has an orientation indicating the dominance of the Tb magnetic subsystem; however, $H_C$ has a higher value than in the as-deposited state. Increasing $D$ to $3\times10^{15}$ He+/cm² (Fig. 2d) results in a modification of the layer system such that the Co magnetic subsystem starts to dominate for Tb layer thicknesses $t_{Tb} \leq 1.6$ nm, i.e. the Tb magnetic subsystem is modified more than the Co one by the $He^+$-ion bombardment. This is an important result, paving the way for an engineering of magnetic patterns without DWs.

To prove such a possibility, we performed local $He^+$ ion bombardment through a resist mask with two doses $D = 1\times10^{15}$ $He^+/cm^2$ and $D = 3\times10^{15}$ $He^+/cm^2$ (see methods and supplementary material) for a selected Tb sublayer thickness of 1.1 nm and studied the magnetization reversal of the magnetically patterned (Tb-1.1nm/Co-0.66nm)$_6$ multilayer. For each dose four 1x1mm² areas were patterned on the same sample with periodically arranged squares of side lengths $a$ = 3, 12.5, 25, and 100 µm and distances between the centers of neighboring squares of $2a$ (see Fig. S1 in supplementary materials). The squares have been modified by ion bombardment, the rest of the sample (matrix) remained unchanged.

Full and minor P-MOKE hysteresis loops for both doses and in all patterned areas are shown in Figs. 3a, 3b, the magnetic moment configurations of the two magnetic subsystems of the ferrimagnet corresponding to the states 1 - 4 observed in the loops are sketched in Figs. 3e and 3f. Additional reference MOKE-measurements were performed on 1×1 mm² square areas bombarded with $D = 1\times10^{15}$ $He^+/cm^2$ and $D = 3\times10^{15}$ $He^+/cm^2$, as well as for an area of the same size, protected by the resist mask (Figs. 3c, 3d). Note that the dimensions of the reference areas were much larger than the laser spot (diameter 0.3 mm) used for MOKE characterization. Therefore, the reference loops are not affected by the border regions between bombarded and not bombarded areas. The situation is different for the patterned periodic square lattices where hysteresis loops are approximately the superposition of the loops obtained for the reference areas. The P-MOKE signal ratio corresponding to magnetization reversal of the squares and matrix is equal to the ratio of the areas of these regions, which is 1/3. Only for the largest squares the observed ratio is not exactly 1/3 as for these measurements the size of the individual squares is close to the MOKE laser spot; in consequence the signal does not fully average over several squares and depends on the precise position of the laser spot with respect to the large squares. A comparably small dependence of the switching fields ($H_S$) on the square size parameter $a$ is observed for switching between states 2→3 and between 4→1, whereas essentially no dependence is observed for the switching between states 3→4 and 1→2 (Figs. 3a and b)). This indicates a relatively weak interaction between the ion-modified regions and the matrix and is caused by weak magnetostatic interactions that result from the low saturation magnetization ($M_S$) of the studied films and their small thicknesses. Exchange coupling at the borders between squares and matrix contributes weakly to the above interaction because of the small interaction surface (the film thickness multiplied by the total perimeter lengths of all the squares).

Magnetization reversal in a RE+ matrix with embedded RE+ squares

The loops corresponding to this case are displayed in Fig. 3a), the magnetic moment configurations of the two magnetic subsystems and the effective magnetizations for the states 1 – 4 are shown in Fig. 3e). Switching fields indicate a dependence on the square dimensions only in the magnetization reversal between states 2→3 and 4→1 (Fig. 3a). Hereinafter, the switching fields $H_S^{if}$, related to the transition between specific states will be described using superscripts identifying the initial ($i$) and final ($f$) states, e.g., for $D = 1\times10^{15}$ $He^+/cm^2$ (Fig. 3a) $H_S^{23}$ and $H_S^{41}$ corresponds to the magnetization reversal of areas (squares) subjected to ion bombardment. At this dose $D$, both the matrix and the squares show the dominance of the magnetic moments of the Tb magnetic subsystem. Therefore, during the transition between states 2→3 and 4→1 the reversals of squares correspond to an annihilation of domains and their corresponding DWs (cf. inset in Fig. 3e). Since the DW energy released by these processes is proportional to the wall interface area, this reduction of $H_S^{23}$ ($H_S^{41}$) with decreasing $a$ is understandable. Additionally, it is obvious that a reduction of $a$ produces a broadening of the transition region for the switching fields

$H_S^{23}$ and $H_S^{41}$. This is related to the statistical variation of $H_S$ among the squares (see movie in supplementary materials). The distribution of switching fields for squares reflects local (lateral) fluctuations of magnetic properties (mainly anisotropy and exchange constants, i.e., parameters determining the energy of DWs) [12]. As the magnetization reversal processes 2→3 and 4→1 correspond to the annihilation of domains and DWs processes 1→2 and 3→4 are related to their creation (Fig. 3e). As the processes 1→2 and 3→4 take place through propagation of a DW in the matrix (in this case the matrix can be treated as a continuous layer [Suppl. Mat]) the magnetization reversal takes place in a very narrow magnetic field range and the values of $H_S^{12}$ and $H_S^{34}$ are equal to the field $H_C$ of the matrix (Fig. 3a, 3c). Moreover, they are practically independent of $a$.

The minor loop shift ($H_{mls}$) (Fig. 3a), measured from the negative saturation field, show positive values for $D = 1 \times 10^{15}$ He$^+$/cm$^2$, revealing a ferromagnetic interaction between the modified areas and the matrix [47]. This is consistent with the tendency to eliminate antiparallel orientations of the magnetization between the magnetic subsystems of the same type (Co and Tb) on opposite sides of the border between squares and matrix [38], i.e. to annihilation of DWs.

The magnetization reversal of the magnetically textured ferrimagnetic films (Fig. 3a), in which RE+ areas (squares) are embedded in a RE+ matrix with different $H_S$, practically does not deviate from a situation in which the ferrimagnetic film would be replaced by a ferromagnetic one. However, the situation changes when the modified areas and the matrix differ not only in $H_S$ but also in magnetic subsystem domination.

Magnetization reversal in a RE+ matrix with embedded TM+ squares

The loops corresponding to this case are displayed in Fig. 3b, the magnetic moment configurations of the two magnetic subsystems and the effective magnetizations for the states 1 – 4 are shown in Fig. 3f. Fig. 3b shows measurements in a system for which the ion bombarded areas have lower $H_S$ and are TM+, while the matrix has a higher $H_S$ and is RE+. In this case, 1→2 and 3→4 magnetization reversals occur in the square areas modified by ion bombardment. In states 1 and 3 (at saturation), the effective magnetizations of squares and matrix are both oriented in the direction of the magnetic field; at the same time, the magnetizations of each magnetic subsystem (Co and Tb) change to the antiparallel direction across the borders between squares and matrix (Figs. 3f, 4g). Therefore, in states 1 and 3, DWs exist at the borders of the squares. At fields $H_S^{12}$ and $H_S^{34}$, the squares reverse (the effective magnetizations of squares and matrix are now antiparallel to each other) and the DWs in the magnetic subsystems are annihilated. To corroborate this conclusion micromagnetic simulations have been carried out to determine the magnetic configuration of the Co and Tb subsystems in the transition area between RE+ and TM+ region. The results are shown in Fig. 4 and will be discussed below.

The comparison of the magnetic configurations of states 1 (3) and 2 (4) (Fig. 3b, 3f) indicates that, at remanence, states 2 and 4 are energetically more favorable than states 1 and 3. This is caused both by a reduction of the magnetostatic energy (the effective magnetization in the squares is antiparallel to that of the matrix) and the annihilation of DWs. As a result, processes 1→2 and 3→4 involve a reduction of the free energy in the system; while the opposite processes are accompanied by an increase (2→1 and 4→3, seen in the minor loops). Although the individual squares reverse independently, the values $H_S^{12}$ and $H_S^{34}$ are close to the $H_C$ value of the modified reference area (Fig. 3b, 3d) and show a narrow distribution, while $|H_S^{21}|$ and $|H_S^{43}|$ (the transition 4→3 is not shown in Fig. 3b) are greater than the $|H_C|$ of the reference area and have large spread (Fig. 3g). The influence of $a$ on the above-mentioned switching fields is stronger for smaller $a$ (or for greater combined length of all DWs). In contrast to the reversal process presented in Fig. 3a, the shift of minor loops seen in Fig. 3b is negative, indicating an antiferromagnetic coupling between the TM+ squares and the RE+ matrix. However, the origin of these behaviors is the same in both cases and it is related to the elimination of the antiparallel configuration of magnetization in the Co and Tb magnetic subsystems (annihilation of DWs). The broadening of the distribution of $H_S^{21}$ and $H_S^{43}$ as $a$ is reduced can be attributed to the spatial distribution of magnetic properties due to deposition and ion bombardment through resist [48–50].

To support qualitatively our interpretation of the experimental data, we have performed micromagnetic simulations using the publicly available OOMMF package [51] without any additional extensions. Details of simulations are described in Methods. A typical full loop and a minor loop for the patterned strip (described in methods) are shown in Fig. 4a. The full loop is a two-step hysteresis with intermediate states similar to those described in the discussion of Fig. 3b. Note that in Fig. 3b the dependence of the P-MOKE signal (strongly dominated by the magnetic Co subsystem) on the magnetic field and in Fig. 4a the one of the effective magnetization is shown. State 1 corresponds to magnetic saturation in negative field where the effective moments in both RE+ and TM+ regions are aligned parallel to the field (Fig. 4f), while for state 2 the effective moment in the bombarded area is opposite to that of the matrix (Fig. 4h). Close-up views of these configurations are shown in Figs. 4e and 4g. These transversal cross-sections show the difference between the two states: in state 1 the effective magnetization is negative everywhere but at the interface the magnetization rotates in each magnetic subsystem(i.e., the DWs are present) while; state 2 does not contain a DW although the two regions have opposite effective magnetizations. These two images support the key finding of our paper, namely that the hybrid RE+/TM+ ferrimagnetic layered system can be patterned by keV He-ion bombardment allowing multi-domains without DWs (stage 1). It is worth noting that, due to the strong antiferromagnetic interaction between the Co and the Tb magnetic subsystems, the spin structure of DWs is similar to the one found in antiferromagnets [52,53].

Having shown that in ferrimagnetic films consisting of TM+ areas embedded in an RE+ matrix the antiparallel configuration of the effective magnetization can exist without DWs at the RE+/TM+ interfaces, now we show that at the field induced transition between states 1 and 2 the reduction in anisotropy energy and exchange energy is accompanied by a reduction of the magnetostatic energy. Overall, the flux closure is achieved with the annihilation of the DW. Figs. 4b-d show the anisotropy, magnetostatic and exchange energies as a function of field for the down sweep branch of the hysteresis loop. In state 2, which occupies the middle region of this graph (-10 kOe < $H$ < -1 kOe), we see that due to annihilation of the DWs the exchange and sum of anisotropy and magnetostatic energies are reduced. It is also apparent that the magnetostatic energy in state 2 is generally lower than in state 1. Therefore, such magnetic configuration is very stable and is characterized by a deep free energy minimum, which explains the strong negative value of $H_{mls}$ observed both in the experiment (Fig. 3b) and simulations (Fig. 4a). This confirms that it is possible to achieve flux closure in the absence of DWs which explains why the observed unique features are particularly stable and energetically advantageous.

## Summary

It has been shown that in a prototypical rare earth (RE)/transition metal (TM) layered ferrimagnetic material system magnetic domains can be engineered by 10 keV He$^+$ ion bombardment without DWs between the patterns. It has been shown that these patterns display a particularly stable magnetic configuration due to a deep minimum in the free energy of the system which is caused by flux closure and the corresponding reduction of the magnetostatic energy part of the total free energy. As a result, a much larger magnetic field is required to annihilate such a magnetic pattern than to create it. The fundamental effect used for engineering of such domains without DWs is the observation that the rare earth contribution to the effective properties of the ferrimagnetic multilayers is more sensitive to keV light-ion bombardment as compared to the contribution of the transition metal. Therefore, this technique can be used in this material system to achieve a steering of the relative contributions of the two magnetic subsystems *in a controlled manner*. Thus, starting with magnetic Co/Tb multilayers where the Tb magnetization dominates and using ion bombardment, we created magnetic patterns where areas with Co magnetic moment density domination and small coercive fields were embedded in the matrix that retained the magnetic properties of the as-deposited system.

**Methods**

**Samples deposition.** The (Tb-wedge 0-2nm/Co-0.66nm)$_6$ and (Tb-1.1nm/Co-0.66nm)$_6$ layered systems were deposited from elemental targets using magnetron sputtering in an ultra-high vacuum chamber (base pressure $10^{-9}$ mbar) with an argon pressure of $10^{-3}$ mbar on 20x20 mm$^2$ naturally oxidized Si(100) substrates coated with a Ti-4 nm/Au-30 nm buffer layer [39]. The wedge-shaped sublayers were produced using a linear shutter. The growth of the films was carried out at RT and, in contrast to our previous investigations [39], without magnetic field. To prevent oxidation of samples an additional 5 nm thick Au protective layer is used.

**Ion bombardment.**

Choice of Tb thickness in Co/Tb multilayers designed for magnetic patterning.

The multilayer Si/Ti-4nm/Au-30nm/(Tb-wedge 0-2nm/Co-0.66nm)$_6$/Au-5nm sample was subjected to two different doses ($D = 1 \times 10^{15}$ and $D = 5 \times 10^{15}$ He$^+$/cm$^2$) of He$^+$ ions. For both $D$ values, a strip of 2mm width was bombarded across the entire sample and parallel to the Tb thickness gradient.

Magnetic patterning

A layered Si/Ti-4nm/Au-30nm/(Tb-1.1nm/Co-0.66nm)$_6$/Au-5nm film was magnetically patterned by bombardment with He$^+$ 10keV ions with two different doses: $D = 1 \times 10^{15}$ He$^+$/cm$^2$ and $D = 3 \times 10^{15}$He$^+$/cm$^2$ [54]. The patterning was carried out by covering the layer system with 400 nm thick photoresist (this thickness is enough to protect the film from ion modification). A mask was used to photolithographically pattern four distinct areas. The patterns in these areas consisted of periodic arrays of squares of side $a = 3, 12.5, 25, 100$ µm, with the centers of neighboring squares separated by a distance of $2a$ Fig. S1 in supplementary materials). The total area of each of these arrays was $1 \times 1$mm$^2$, i.e. large enough for hysteresis loops measurements using our P-MOKE magnetometer (which has a laser spot of 0.3 mm in diameter). Independently, a $1 \times 1$mm$^2$ area not covered with the photoresist was also manufactured for reference. The above described pattern was replicated for experiments with different values of $D$.

**Magnetic measurements.** Magnetooptical hysteresis loops in polar configuration (P-MOKE) were measured in the same way as described in our previous paper [39] using a laser with 640 nm wavelength. Images of magnetic structure and movies illustrating magnetization reversal process were recorded using a P-MOKE microscope.

**Micromagnetic simulations**

To simulate the ferrimagnetic alloy film, we used cubic discretization cells with very small size (1nm). For each cell a uniform random number has been assigned which determines the material of which it is made. With this procedure, the alloy is modelled as a cubic granular structure with random occupancy by the RE element. We believe that the granular structure captures two important physical features: first, because the individual sublayers of Co/Tb multilayer are very thin the system does not form continuous films but tends to behave as an alloy; second, the difference in atomic sizes between the RE and the TM causes the structure to be amorphous rather than crystalline. In this way, the granular structure used in our simulations resembles the formation of islands during the deposition procedure. Using this approach, two features of ferrimagnets at compensation can be demonstrated qualitatively: the vanishing of magnetization saturation and the unbounded growth of the coercive field.

We emphasize that the cited parameter values used in the micromagnetic modelling are given solely to facilitate replication of our micromagnetic simulations. We do not claim to have obtained a quantitative agreement between simulations and experiment. We show instead that the qualitative features can be reproduced in the simulation. A quantitative matching would require performing numerical analysis of the errors introduced when a continuous alloy is represented by discrete grains. This task is beyond the scope of this paper. For this reason, in this micromagnetic simulation section we would refer to the different elements in the structure using generic names (RE and TM).

The effect of ion bombardment in the ferrimagnet is modelled by separating the system into two distinct regions. Inside the bombarded area we reduce the RE occupancy probability and decrease the strength of the crystalline anisotropy of the TM. We simulate a strip long enough in one direction to cover a full period of the structure. The simulation box is 4μm×20nm×5nm with periodic boundary conditions in two dimensions. Longitudinally, the bombarded area is placed in the central region with margins of 1μm from each end of the strip; in the transversal and vertical directions it spans the whole simulation box.

To describe RE+ and TM+ regions (corresponding, in the experiment, to protected and bombarded areas, respectively) four material regions are used to specify: first, whether the cell is occupied with RE or with TM elements; and second, if the cell is in the pristine (out) or the bombarded area (in). Any cell in the simulation belongs to one of the following regions: $RE_{in}$, $RE_{out}$, $TM_{in}$, $TM_{out}$. The parameters were chosen to capture the following known properties of ferrimagnets [55,56]: weak ferromagnetic interaction between neighboring RE-RE cells, a stronger ferromagnetic interaction between neighboring Co-Co atoms, and an even stronger antiferromagnetic interaction between adjacent TM-RE cell pairs. The magnetic moment of the RE cells was chosen to produce a compensation point at 22% [57]. The easy-axis of effective anisotropy is oriented perpendicular to the surface. The crystalline anisotropy constant for the TM is weaker for cells located in the bombarded area but is everywhere larger than that of RE cells which all have the same value assigned. The material parameters are summarized in Table 1

| Region (x) | $K_u \left( \frac{MJ}{m^3} \right)$ | $M_S \left( \frac{MA}{m} \right)$ | $A_{x-RE} \left( \frac{pJ}{m} \right)$ | $A_{x-TM} \left( \frac{pJ}{m} \right)$ |
|---|---|---|---|---|
| $RE_{in}$ | 1.067 | 5.08 | 7 | -24 |
| $RE_{out}$ | 1.067 | | | |
| $TM_{in}$ | 1.342 | 1.42 | -24 | 14 |
| $TM_{out}$ | 1.742 | | | |

*Table 1 Material Parameters. The exchange constants $A_{x-y}$ should be read as the coupling between elements in row x and column y. These ad-hoc values have been chosen to reproduce the qualitative features of our experiments and should not be considered as the estimations of actual material parameters.*

**Acknowledgements**
The authors would like to thank M. Schmidt and J. Aleksiejew for technical support. The work was financed by the National Science Centre Poland under SONATA BIS funding UMO-2015/18/E/ST3/00557. M.M. and A.M. acknowledge financial support from the National Science Centre Poland through the SONATINA project UMO-2018/28/C/ST5/00308.

**References**
[1]    A. Moser, K. Takano, D. T. Margulies, M. Albrecht, Y. Sonobe, Y. Ikeda, S. Sun, and E. E. Fullerton, J. Phys. Appl. Phys. **35**, R157 (2002).

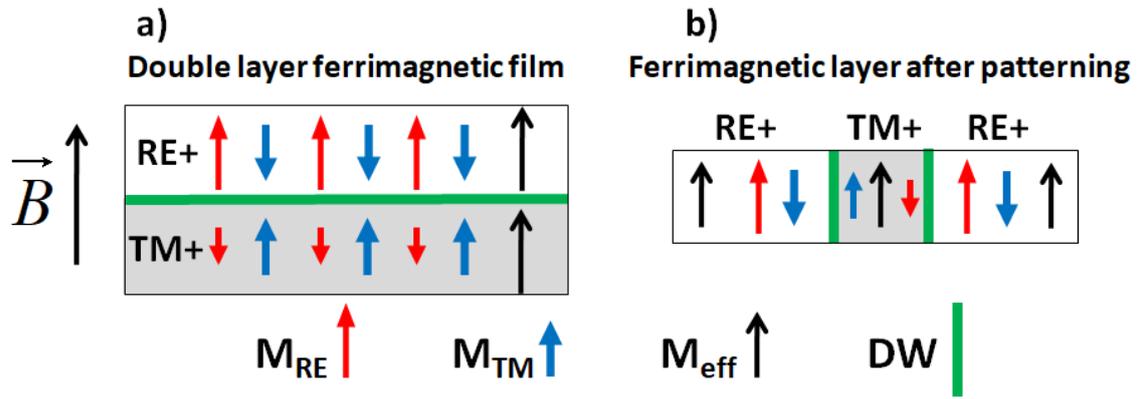

Fig. 1 a) Sketch of a layer system consisting of a stack of an RE+ and a TM+ ferromagnetic film. Red and blue arrows indicate the magnetizations of the RE and TM magnetic subsystems, respectively, black arrows indicate the effective magnetization of the layers. The magnetization configuration is depicted at magnetic saturation, displaying an interfacial DW (green area) between the RE+ and the TM+ layer b) Sketch of a ferrimagnetic layer, displaying alternating RE+ and TM+ regions after local modification by keV light-ion bombardment. The magnetic configuration is also depicted at magnetic saturation, displaying DWs (green areas) between the RE+ and TM+ regions.

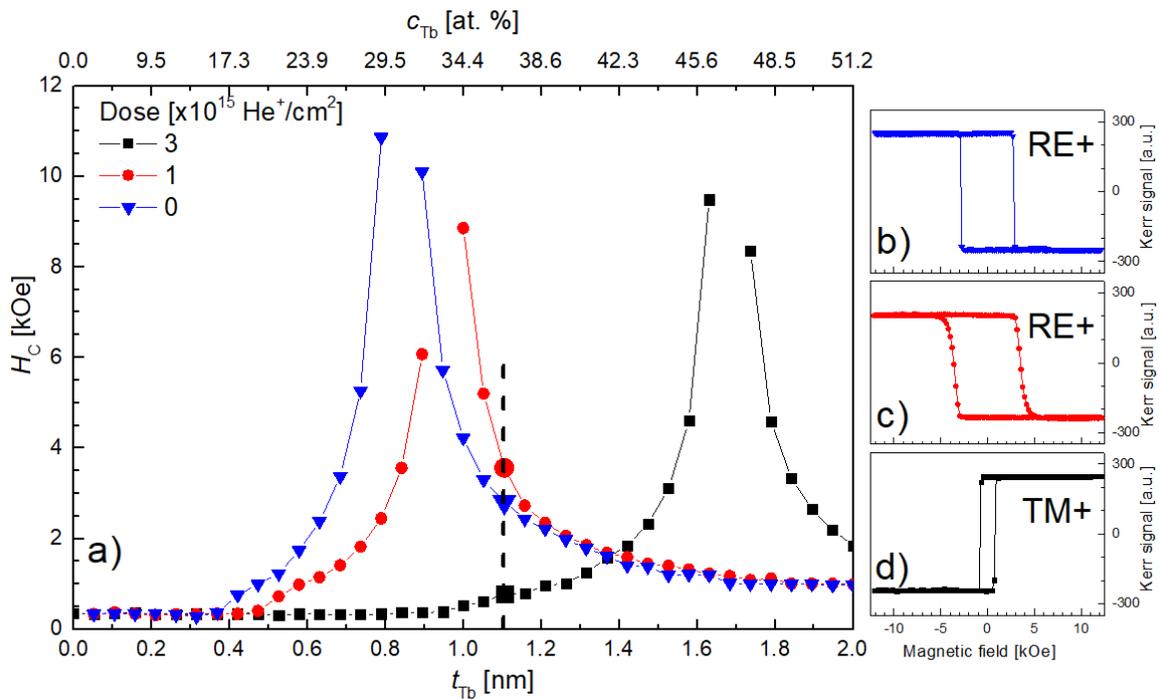

Fig. 2 a) Coercive field $H_C$ as a function of Tb sublayer thicknesses of the Si/SiOx/Ti-4nm/Au-30nm/(Tb-wedge/Co-0.66nm)$_6$/Au system in the as-deposited state and after He$^+$ (10 keV) ion bombardment with $D=1\times10^{15}$ He$^+$/cm$^2$ and $D=3\times10^{15}$ He$^+$/cm$^2$. The upper horizontal axis shows the corresponding effective concentration of Tb in the whole layer system, $c_{Tb}$, for a given Tb thickness. The dashed line corresponds to $t_{Tb} = 1.1$ nm which was chosen for experiments presented in Fig. 3. The hysteresis loops corresponding to large points ($t_{Tb}=1.1$ nm) in panel a) are presented in panels b), c) and d) for $D=0$, $D=1\times10^{15}$He$^+$/cm$^2$, and $D=3\times10^{15}$He$^+$/cm$^2$, respectively.

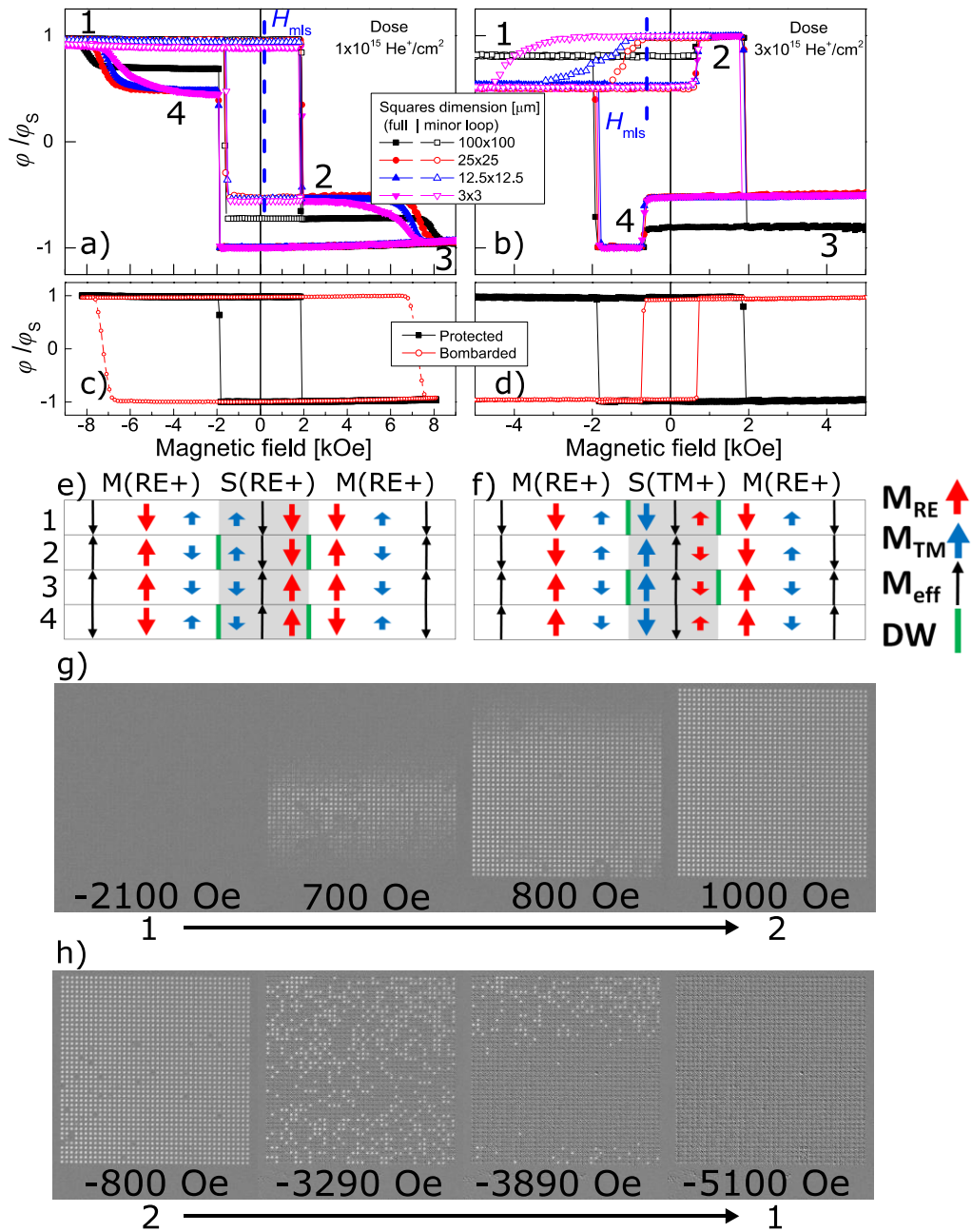

Fig. 3 Full and minor (full and open symbols, respectively) P-MOKE hysteresis loops measured for a Si/Ti-4nm/Au-30nm/(Tb-1.1nm/Co-0.66nm/)$_6$/Au-5nm system magnetically patterned using ion bombardment (He$^+$ 10 keV) with doses $D = 1\times10^{15}$ He$^+$/cm$^2$ (a,c) and $D = 3\times10^{15}$ He$^+$/cm$^2$ (b,d). The different colors in panels (a, b) correspond to different sizes of patterned squares. The hysteresis loops presented in the lower panels correspond to reference areas (c, d). The magnetic field corresponding to the minor loop shift $H_{mls}$ is indicated only for a=12.5 μm. The panels (e, f) show the magnetization orientation in the matrix (M) and the squares (S). The black, blue and red arrows correspond to effective magnetization, magnetization of the Co and of the Tb magnetic subsystems, respectively. DWs are indicated with green. The magnetic structure inside the DW is shown in Fig.4e. Panels (g,h) show differential images (difference between images recorded at a given magnetic field and at saturation in negative field) of magnetic structure recorded using magnetooptical Kerr microscope in polar configuration. The photographs are arranged in rows corresponding to magnetic field ranges related to the minor loop reversal of the 12.5×12.5 μm squares from 1 to 2 g) and from 2 to 1 h).

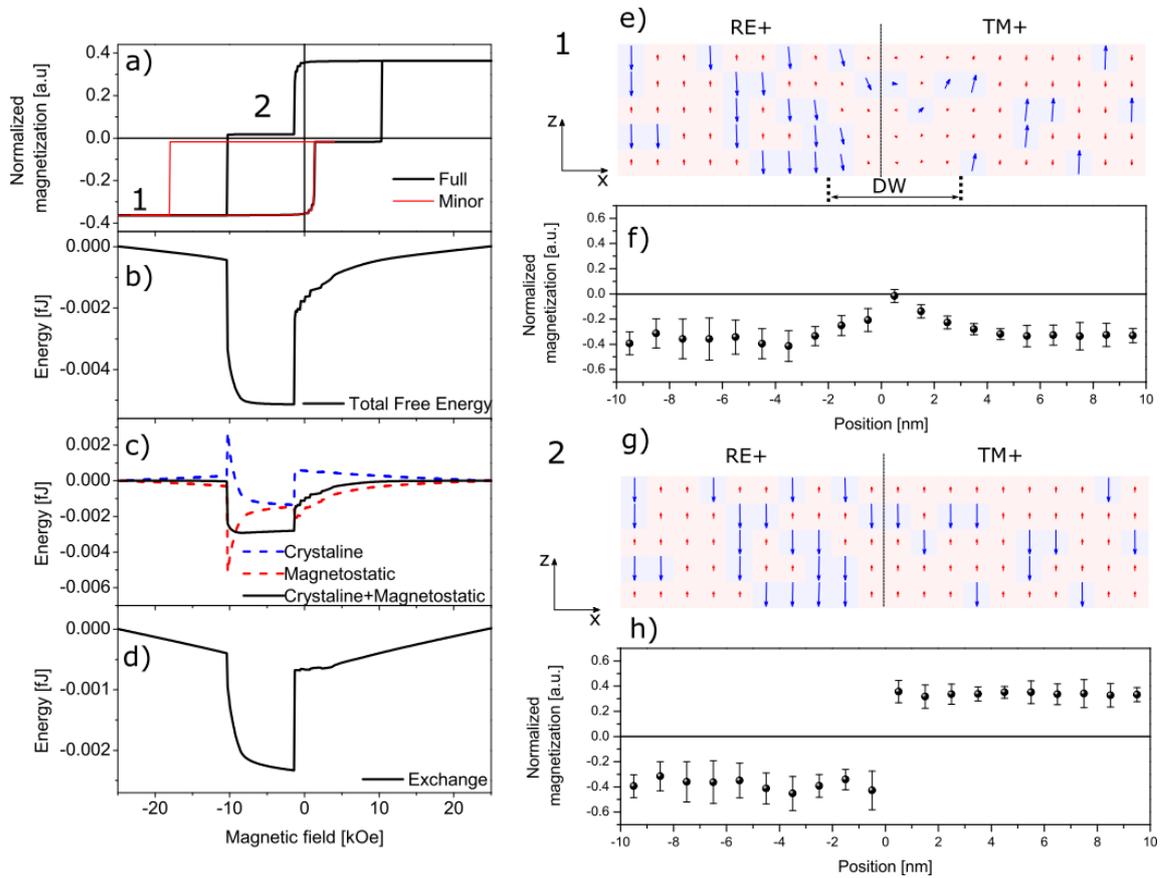

Fig. 4 a) Hysteresis loops obtained from OOMMF simulations for a patterned strip for randomized distributions of Tb cells. The magnetization perpendicular to the plane, $m_z$, is normalized accounting for the total number of Co and Tb cells. b) Free energy (magnetostatic+anisotropy+exchange), c) Sum of anisotropy and magnetostatic energies; and d) exchange energy as functions of applied field for the sweep of the hysteresis loop from 25 kOe to -25 kOe. To facilitate comparison, the energy terms in the saturated state are set to zero. (e, g) Cross section of the Co (red arrows) and Tb (blue arrows) magnetization configuration in the region between RE+ and TM+ areas at magnetic field $H = 25$ kOe and $H = -3$ kOe corresponding to state 1 and 2. (f, h) Normalized $m_z$ component at distance x away from the boundary between the RE+ and the TM+ regions. In state 1 (at saturation) a DW is present - the Co and Tb spins rotate along the x-direction with continuous changes in normalized magnetization keeping its sign (f). In contrast, state 2 shows no DW between magnetic domains with antiparallel magnetization (h). The error bars in f) and h) correspond to the standard deviation of 11 simulations with different random distributions.